\documentclass[aps,prd,twocolumn,floatfix,nofootinbib,showpacs,superscriptaddress,tightenlines]{revtex4}

\usepackage{amsmath}
\usepackage{lipsum}
\usepackage{amssymb}
\usepackage{amsthm}
\usepackage{bbold}
\usepackage{dcolumn}
\usepackage{epsfig}
\usepackage{graphics}
\usepackage{graphicx}
\usepackage{longtable}
\usepackage{color}

\usepackage{xspace}
\usepackage{cancel}

\newcommand{\me}{\mathrm{e}} 
\newcommand{\mi}{\mathrm{i}} 

\newcommand{\eqn}[1]{Eq.~(\ref{#1})}

\newcommand{\dfd}[2]{\ensuremath{\frac{\mathrm{d}^{#2}#1}{(2\pi)^{#2}}}\,}
\newcommand{\dfdinv}[2]{\ensuremath{\mathrm{d}^{#2}#1}\,}
\newcommand{\dx}[1]{\ensuremath{\mathrm{d}#1\,}}

\newcommand{\nn}{\nonumber}

\begin{document}

\title{Landau-Khalatnikov-Fradkin transformations in reduced quantum electrodynamics}

\author{A.~Ahmad$^{1,2}$, J.~J.~Cobos-Mart\'{\i}nez$^1$, Y.~Concha-S\'anchez$^3$ and
A.~Raya}

\affiliation{
Instituto de F\'{i}sica y Matem\'aticas, Universidad Michoacana de San
Nicol\'as Hidalgo, Edificio C-3, Ciudad Universitaria, 
Morelia, Michoac\'an 58040, Mexico\\
$^2$ Department of Physics, Gomal University, 29220 D.I. Khan, K.P.K., Pakistan.\\
$^3$Facultad de Ingenier\'{i}a Civil, Edificio C, Ciudad Universitaria, 
Morelia, Michoac\'an 58040, Mexico.
}

\date{\today}

\begin{abstract}

We derive the Landau-Khalatnikov-Frandkin transformation (LKFT) for the fermion propagator in quantum electrodynamics (QED) described within a brane-world
inspired framework where photons are allowed to move in $d_\gamma$ space-time (bulk)  dimensions, while electrons remain confined to a $d_e$-dimensional brane,
with $d_e < d_\gamma$, referred to in the literature as reduced quantum electrodynamics, RQED$_{d_\gamma,d_e}$. Specializing to the case of graphene, 
namely, RQED$_{4,3}$ with massless fermions, we derive the nonperturbative form of the fermion propagator starting from its bare counterpart and then compare 
its weak coupling expansion to known one- and two-loop perturbative results. The agreement of the gauge-dependent terms of order $\alpha$ and $\alpha^{2}$ is
reminiscent from the structure of LKFT in ordinary QED in arbitrary space-time dimensions and provides strong constraints for the multiplicative renormalizability of RQED$_{d_\gamma,d_e}$.

\end{abstract}


\maketitle

\date{\today}

\section{\label{sec:intro}Introduction}

Gauge symmetry is the cornerstone of our current understanding of the fundamental interactions among the building blocks of the Universe. 
Quantum electrodynamics (QED) is probably the best-known example of a quantum field theory with an underlying gauge symmetry where the 
theoretical predictions (based upon its the multiplicative renormalizability character) and the experimental results meet with remarkable agreement; 
for example, the anomalous magnetic moment of the muon is in agreement with 
the experimental 
value up
to 
six significant 
digits~\cite{Marquard:2015dra,Aoyama:2012wj}. The gauge principle in QED at the level of the corresponding Green functions is reflected in sets of 
relations among different $n$-point functions. Ward~\cite{ward},  Ward-Green-Takahashi~\cite{ward,green,takahashi}, and transverse
Ward-identities~\cite{twi1,twi2,twi3,twi4,twi5} relate $(n+1)$-point to $n$-point functions in constructions resembling divergence and curl of currents, 
while Nielsen identities~\cite{ni1,ni2} guarantee the gauge invariance of poles of propagators at one loop~\cite{ni3} and to all orders 
in perturbation theory~\cite{ni4,ni5}. A different family of transformations dealing with the gauge covariant character of QED is the 
Landau-Khalatnikov-Fradkin transformations (LKFT)~\cite{Landau:1955zz,Fradkin:1955jr}, which describe in coordinate space the specific manner in which 
a given Green function, either perturbative or nonperturbative in nature, transforms covariantly in different gauges. These transformations have been 
derived by different authors and different approaches in the past decades~\cite{LKF1,LKF2,LKF3,LKF4,LKF5}. For the fermion propagator, these transformations 
have been extensively used to establish multiplicative renormalizability of the theory, by imposing perturbative constraints on the 
charged-particle-photon vertex in spinless~\cite{Fernandez-Rangel:2016zac,Ahmadiniaz:2015kfq} and spinor QED~\cite{Kizilersu:2009kg,Bashir:2001vi}. 
The nonperturbative nature of the LKFT allows us to fix some of the coefficients of the all-order expansion of the fermion propagator. Starting with a perturbative
propagator at a fixed order $n_o$ in perturbation theory in Landau gauge, all the coefficients dependent of the gauge parameter of the propagator at
order $(n_o+1)$ get fixed by the weak coupling expansion of the LKF-transformed initial one. The LKF transformation for the fermion propagator has been
extensively used in three-dimensional QED (QED$_3$)~\cite{Bashir:2002sp,Bashir:2004rg,Bashir:2008fk,Bashir:2000,Bashir:2008ej,Bashir:2005wt,Bashir:2009} and more recently extended to
QCD~\cite{Aslam:2015nia,Aguilar:2015nqa}. In the particular case of QED$_3$ --which is regarded as an effective model of high-energy, large fermion 
family number approximation to QCD--the LKFT allows a direct description in momentum space and hence has been widely implemented to address gauge-invariant
issues
in
nonperturbative
studies
of
dynamical
chiral
symmetry
breaking
and
confinement within the Schwinger-Dyson equations
framework~\cite{Bashir:2008fk,Sauli:2010kk,Fischer:2004nq,Burden:1993gy}. 
QED$_3$ has also been traditionally used to describe a number of condensed matter systems, 
including quantum Hall effect systems~\cite{Fialkovsky:2011wh}, high-Tc superconductors~\cite{Herbut:2002yq}, and more recently
graphene~\cite{Fialkovsky:2011wh,Cortijo:2011aa} and other Weyl semimetals~\cite{Miransky:2015ava}.

The new era of materials science emerging after graphene has opened new avenues to explore applying
ideas in the particle physics realm to condensed
matter systems. Dirac-Weyl semimetals are a class of crystals which have conic dispersion relations near the Dirac points in the first Brillouin zone in such
a way that the charge carriers, which are confined to two-dimensional membranes,  are described by an effective low-dimensional Dirac equation, while the
electromagnetic field quanta move unconstrained throughout space. Such dynamics resembles brane-world scenarios where the photon plays the role of the 
graviton and is allowed to move in bulk dimensions $d_\gamma$, while the matter fields--electrons--are restricted to have dynamics on a $d_e$-dimensional 
brane with $d_e<d_\gamma$. The framework describing this scenario has been dubbed reduced QED (RQED$_{d_\gamma,d_e}$)~\cite{Gorbar:2001qt}. The particular case 
of RQED$_{4,3}$ with massless fermions is regarded as the physical realization of low-energy graphene and other Weyl-Dirac systems. Hence the importance 
of investigating the gauge covariance properties of the fermion propagator in RQED$_{d_\gamma,d_e}$.

The multiplicative renormalizability character of RQED$_{d_\gamma,d_e}$ for massless fermions has been verified up to the two-loop order in 
Refs.~\cite{Kotikov:2013eha,Teber:2014hna,Teber:2012de}. Here we verify this statement by investigating the gauge covariance properties of the 
fermion propagator in RQED$_{4,3}$ through the corresponding LKFT. We adapt the successful strategy implemented in 
Refs.~\cite{Bashir:2002sp,Ahmadiniaz:2015kfq,Aslam:2015nia,Bashir:2000,Bashir:2008ej,Bashir:2005wt}, starting with the fermion propagator at tree level in Landau gauge and 
LKF-transform it nonperturbatively to other gauges. Then, we perform a weak coupling expansion of our findings and compare against the perturbative
results of Ref.~\cite{Kotikov:2013eha} allowed by the structure of LKFT. For that purpose, we have organized the remaining of this article as follows: in Sec.~\ref{QED} we review 
the fermion propagator in the light of LKFT in ordinary QED in arbitrary space-time dimension $d$ but specialize in the case $d=3$. We briefly describe  RQED$_{d_\gamma,d_e}$ and 
derive the corresponding LKFT in Sec.~\ref{RQED}. Perturbative constraints of the structure of the fermion propagator in RQED$_{4,3}$ are discussed 
in Sec.~\ref{pert}. We conclude in Sec.~\ref{conclusions} and present some auxiliary integrals in the Appendix.

\section{Fermion propagator in QED}~\label{QED}

We start our discussion by considering the general structure of the Dirac fermion propagator in QED. In momentum space, the fermion two-point function 
$S(p;\xi)$ has the general form 

\begin{equation}
\label{eqn:SpSpace}
S(p;\xi)=-\frac{F(p;\xi)}{p^{2}+M^{2}(p;\xi)}\left(\mi\cancel{p}+M(p;\xi)\right),
\end{equation}

\noindent where $F(p;\xi)$ is the so-called wave function renormalization and $M(p;\xi)$ is the fermion mass function. We have included the gauge parameter
$\xi$ dependence of these functions because we are interested in the form of the propagator in different covariant gauges. On the other hand, in coordinate 
space, $S(x;\xi)$ can generally be written as

\begin{equation}
\label{eqn:SxSpace}
S(x;\xi)=\cancel{x}X(x;\xi) + Y(x;\xi).
\end{equation}

\noindent Equations (\ref{eqn:SpSpace}) and (\ref{eqn:SxSpace}) are valid for any space-time dimensionality $d$ and related to each other by a Fourier
transformation, namely,

\begin{eqnarray}
\label{eqn:FT}
S(p;\xi)&=&\int\dfdinv{x}{d}\me^{\mi p\cdot x}S(x;\xi), \\
\label{eqn:InvFT}
S(x;\xi)&=&\int\dfd{p}{d}\me^{-\mi p\cdot x}S(p;\xi).
\end{eqnarray}

\noindent Correspondingly, in momentum space, the free gauge boson propagator $D_{\mu\nu}(p)$ takes the form

\begin{equation}
\label{eqn:photond}
D_{\mu\nu}(p)=\frac{-i}{p^2} \left(g_{\mu\nu}-\frac{p_\mu p_\nu}{p^2}\right)+\xi\frac{p_\mu p_\nu}{(p^2)^2}\,, 
\end{equation}

\noindent in any number of space-time dimensions. The longitudinal part of this propagator, proportional to the gauge parameter $\xi$ ($\xi$=0 corresponds to 
the Landau gauge) and inversely proportional to $p^4$, points  to the specific manner in which this two-point function varies from gauge to gauge 
and is crucial to the derivation of the LKFT for the fermion
propagator~\cite{Landau:1955zz,Fradkin:1955jr,LKF1,LKF2,LKF3,LKF4,LKF5}. This transformation is more clearly written in coordinate space and states that the
fermion propagator in an arbitrary covariant gauge $S(x;\xi)$ is related to the corresponding propagator in Landau gauge $S(x;0)$ through the 
transformation

%
%

\begin{equation}
\label{eqn:LKFTQEDd}
S_{d}(x;\xi)=S_{d}(x;0)\me^{-\mi[\Delta_{d}(0)-\Delta_{d}(x)]}.
\end{equation}

\noindent The function $\Delta_{d}(x)$,  which essentially defines the LKF transformation \eqn{eqn:LKFTQEDd}, is defined as~\cite{Landau:1955zz,Fradkin:1955jr}

\begin{equation}
\label{eqn:DeltadQED}
\Delta_{d}(x)=-\mi\xi e^{2}\mu^{4-d}\int\dfd{q}{d}\frac{\me^{-\mi q\cdot x}}{q^{4}},
\end{equation}

\noindent where $e$ is the fermion electric charge, and $\mu$ is a mass scale introduced such that $e$ is dimensionless in four dimensions, but yields a dimensionful coupling $\alpha=e^2/(4\pi)$ in QED$_3$. 
Equation~(\ref{eqn:DeltadQED}) is related to the Fourier transform of the longitudinal part of the gauge boson propagator~\cite{Landau:1955zz,
Fradkin:1955jr}, and the momentum integration is over the gauge boson momentum. 
Performing the required integrations, $\Delta_{d}(x)$, is explicitly given by~\cite{Bashir:2002sp}

\begin{equation}
\label{eqn:DeltadQEDExpl}
\Delta_{d}(x)=-\frac{\mi\xi\alpha}{4\pi^{\frac{d-2}{2}}}\Gamma\left(\frac{d-4}{2}\right)
(\mu x)^{4-d},
\end{equation}

\noindent where $\alpha=e^2/(4\pi)$ is the coupling constant, and $\Gamma(z)$ is the Euler Gamma function.

The LKFT for the fermion propagator has been widely studied~\cite{Ahmadiniaz:2015kfq,Bashir:2008ej,Bashir:2005wt,Bashir:2002sp}, in particular in QED in three 
and four dimensions for massive and massless fermions. The typical strategy to explore the structure of the fermion propagator through the LKFT works as
follows~\cite{Ahmadiniaz:2015kfq,Bashir:2008ej,Bashir:2005wt,Bashir:2002sp}: to obtain the fermion propagator in any gauge from \eqn{eqn:LKFTQEDd} we provide 
the fermion propagator in a particular gauge, usually Landau gauge. This is most easily done in coordinate space. After this, we  Fourier transform with
\eqn{eqn:FT} to obtain the fermion propagator in momentum space. However,  knowledge of the full fermion propagator $S$ even in a particular gauge 
is formidable. Nevertheless, being nonperturbative in nature, the LKFT actually  provides valuable information on the structure of the fermion propagator: we 
can rely on perturbation theory to provide the starting point $S(x;0)$ or $S(p;0)$, although this has 
some caveats; see Ref.~\cite{Bashir:2002sp}. Nonetheless, we take $F(p;0)$ and $M(p;0)$ as given by the lowest order of perturbation theory:

\begin{eqnarray}
\label{eqn:S0pSpace}
F(p;0)&=&1, \nn \\
M(p;0)&=&m,
\end{eqnarray}

\noindent where $m$ is the current fermion mass. In the relevant case $d=3$, for massless fermions, the LKFT strategy reveals that the nonperturbative fermion
propagator in an arbitrary covariant gauge is (see, for instance, Ref.~\cite{Bashir:2002sp})

\begin{eqnarray}
\label{eqn:SLKF3d}
F(p;\xi)&=&1-\frac{\alpha\xi}{2p}\arctan{\left(\frac{2p}{\alpha\xi} \right)}.
\end{eqnarray}

\noindent Thus, a weak coupling expansion reveals that all terms of the form $(\alpha\xi)^j$ are fixed from Eq.~(\ref{eqn:SLKF3d}). This is a major asset of 
the LKFT. Below we shall derive the corresponding transformation for RQED$_{d_\gamma,d_e}$, and in particular for $d_\gamma=4,d_e=3$.


\section{The LKF transformation for reduced QED}~\label{RQED}

Reduced QED for massless fermions is described from the action~\cite{Gorbar:2001qt,Teber:2012de}

\begin{equation}
\label{eqn:IRQED}
I_{d_\gamma,d_e}[A_{\mu_\gamma},\psi_{(d_e)}]=\int d^{d_\gamma}x\ {\cal L}_{d_\gamma,d_e},
\end{equation}

\noindent where the Lagrangian

\begin{eqnarray}
\label{eqn:LRQED}
{\cal L}_{d_\gamma,d_e}&=& \bar\psi(x) i\gamma^{\mu_e} D_{\mu_e}\psi(x)\delta^{(d_\gamma-d_e)}(x)
-\frac{1}{4}F_{\mu_\gamma \nu_\gamma}F^{\mu_\gamma \nu_\gamma}\nonumber\\ &&-\frac{1}{2\xi}(\partial_{\mu_\gamma}A^{\mu_\gamma})^2
\end{eqnarray}

\noindent includes matter fields $\psi(x)$ restricted to a $d_e$-dimensional brane ($\mu_{e}=0,1,\dots,d_{e}-1$) and gauge fields $A_{\mu_\gamma}(x)$ propagating in $d_\gamma$-bulk 
dimensions ($\mu_{\gamma}=0,1,\dots,d_{\gamma}-1$), with $d_\gamma>d_e$. Here $D_{\mu}$ represents the covariant derivative and $F_{\mu\nu}$ the field strength tensor. The free photon propagator along bulk dimensions is of the same form as in \eqn{eqn:photond}, namely, 

\begin{equation}
\label{eqn:photonbulk}
D_{\mu_\gamma \nu_\gamma}(p)=\frac{-i}{p^2} \left(g_{\mu_\gamma \nu_\gamma}-\frac{p_{\mu_\gamma} p_{\nu_\gamma}}{p^2}\right)+\xi\frac{p_{\mu_\gamma} p_{\nu_\gamma}}{(p^2)^2}\,, 
\end{equation}

\noindent but when reduced to the $d_e$-dimensional brane, it becomes~\cite{Teber:2012de,Teber:2014hna}

\begin{equation}
\label{eqn:photondebrane}
D_{{\mu_e}{\nu_e}}(p)= D(p^2)\left(g_{{\mu_e}{\nu_e}}-\frac{p_{\mu_e} p_{\nu_e}}{p^2}\right)+\tilde{\xi} D(p^2)\frac{p_{\mu_e} p_{\nu_e}}{p^2}. 
\end{equation}

\noindent  Here,

\begin{eqnarray}
D(p^2)& =& \frac{i}{(4\pi)^{\varepsilon_e}}\frac{\Gamma(1-\varepsilon_e)}{(-p^2)^{1-\varepsilon_e}},
\end{eqnarray}

\noindent where $\varepsilon_{e}=(d_{\gamma}-d_{e})/2$ and $\tilde{\xi}=(1-\varepsilon_e)\xi$.
The longitudinal piece of the propagator changes the form of the LKFT for the fermion propagator.
Considering that the propagator changes from gauge to gauge according to

\begin{equation}
\label{eqn:LKFTGraphene}
S_{d_e}(x;\xi)=S_{d_e}(x;0)\me^{-\mi[\tilde{\Delta}_{d_e}(0;\varepsilon_{e})-\tilde{\Delta}_{d_e}(x;\varepsilon_{e})]},
\end{equation}

\noindent we define the function

\begin{eqnarray}
\label{eqn:DeltaRQED}
\tilde{\Delta}_{d_{e}}(x;\varepsilon_{e})&=&-\mi f(\varepsilon_{e})\xi e^{2}\mu^{4-d_{\gamma}}\int\dfd{q}{d_{e}} 
\frac{\me^{-\mi q\cdot x}}{q^{4-2\varepsilon_{e}}} \\
&=&
\label{eqn:DeltaRQEDl2}
-\mi f(\varepsilon_{e})\xi e^{2}\frac{\Gamma(\frac{d_{e}-a}{2})}{2^{a}\pi^{d_{e}/2}\Gamma(\frac{a}{2})}(\mu x)^{a-d_{e}},\nonumber\\
\end{eqnarray}

\noindent where $f(\varepsilon_{e})=\Gamma(1-\varepsilon_{e})(1-\varepsilon_{e})/(4\pi)^{\varepsilon_{e}}$ and $a=4-2\varepsilon_{e}$. This is the 
general form of LKFT for the fermion propagator in RQED$_{d_\gamma,d_e}$ and guarantees that $e^2$ is dimensionless in RQED$_{4,d_e}$~\cite{Teber:2012de,Kotikov:2013eha}. Note that when $d_{\gamma}=d_{e}=d$ we have $\varepsilon_{e}=0$,
$f(\varepsilon_{e})=1$, and~\eqn{eqn:DeltaRQED} reduces to~\eqn{eqn:DeltadQED}, thus recovering the usual LKF transformation for QED~\eqn{eqn:LKFTQEDd} 
in any dimension $d$. Furthermore, in order to obtain \eqn{eqn:DeltaRQEDl2}, we have used the fact  that $2\varepsilon_e+d_e=d_\gamma$ in the mass dimensions of $\mu$. 

%
%
%
%
In graphene ($d_{\gamma}=4$, $d_{e}=3$, $\varepsilon_{e}=1/2$), fermions are massless and move on a plane,  while the photon lives in the usual four-dimensional 
space-time.
Note that in~\eqn{eqn:DeltaRQED}, the power of $q$ in the denominator of the integrand is 3. Furthermore, note that the covariant gauge parameter has been
``reduced'' by a factor of 4 (f($\varepsilon_{e}=\frac{1}{2}$)=1/4). Both of these modifications are a consequence of integrating out the bulk  degrees of
freedom~\cite{Gorbar:2001qt} in \eqn{eqn:DeltadQEDExpl}. Thus, we explicitly find that the function defining the LKFT for the fermion propagator in graphene is
%

\begin{equation}
\label{eqn:DeltaGrapheneExpl}
\tilde{\Delta}_{3}\left(x,\frac{1}{2}\right)=\frac{-\mi\xi e^{2}}{16\pi^{2}}
\Gamma\left(\frac{1-2\epsilon}{2}\right)(\mu x)^{2\epsilon-1},
\end{equation}

\noindent with $\epsilon\to 1/2$. Expanding \eqn{eqn:DeltaGrapheneExpl} around $\epsilon=1/2$, defining $\delta=\epsilon-1/2$ and making use of the expansions

\begin{eqnarray}
\label{eqn:epsilonexpansion1}
a^{x}&=&1+ x\ln(a) + \mathcal{O}(x^{2}), \\
\label{eqn:epsilonexpansion2}
\Gamma(x)&=&\frac{1}{x}-\gamma_{E}+\frac{1}{12}(6\gamma_{E}^{2}+\pi^{2})x
+ \mathcal{O}(x^{2}),
\end{eqnarray}

\noindent $\gamma_{E}$ representing the Euler-Mascheroni constant, we get

\begin{equation}
\label{eqn:DeltaGrapheneExplExp}
\tilde{\Delta}_{3}\left(x,\frac{1}{2}\right)=\frac{\mi\xi e^{2}}{16\pi^{2}}\left[
\frac{1}{\delta}+ \gamma_{E}+2\ln(\mu x)
+ \mathcal{O}(\delta)\right].
\end{equation}


\noindent  Since the transformation function, \eqn{eqn:DeltaGrapheneExplExp}, cannot be evaluated at $x=0$, we introduce a cutoff $x_{min}$, such that 

\begin{equation}
\label{eqn:DeltaGrapheneFinal}
-\mi\left[\tilde{\Delta}_{3}\left(x_{min},\frac{1}{2}\right)-\tilde{\Delta}_{3}\left(x,\frac{1}{2}\right)\right]
= \ln\left(\frac{x}{x_{min}}\right)^{-2\nu},
\end{equation}

\noindent where we have defined $\nu=\xi\alpha/(4\pi)$, and the dimensionless coupling constant $\alpha= e^{2}/(4\pi)$.

With \eqn{eqn:DeltaGrapheneFinal} at hand we are now in a position to compute the fermion propagator in graphene for any gauge from \eqn{eqn:LKFTGraphene}.
For massless fermions  $Y(x;0)=0$, and therefore $Y(x;\xi)=0$ for any covariant gauge. Only $X(x,\xi)$ is nonzero. It is given by

\begin{eqnarray}
\label{eqn:XAnyGaugeGraphene}
X(x;\xi)&=&X(x;0)\me^{-\mi\left[\tilde{\Delta}_{3}\left(x_{min},\frac{1}{2}\right)-\tilde{\Delta}_{3}\left(x,\frac{1}{2}\right)\right]} \nn \\
&=&-\frac{x_{min}^{2\nu}}{4\pi}x^{-2\nu-3}.
\end{eqnarray}


\noindent Furthermore, since in the massless limit $Y(x;\xi)=0$, then $M(p;\xi)=0$ in any covariant gauge. This is consistent with the well-known fact that 
fermion masses cannot be radiatively generated in QED. In this limit, the wave function renormalization is given by

\begin{equation}
\label{eqn:FAnyGaugeGraphene}
-\mi F(p;\xi)=\int\dfdinv{x}{3}(p\cdot x)\me^{\mi p\cdot x} X(x;\xi).
\end{equation}

\noindent Using the formulas given in the Appendix, the wave function renormalization for the fermion propagator is explicitly given by

\begin{equation}
\label{eqn:FAnyGaugeGrapheneExpl}
F(p;\xi)=\frac{\sqrt{\pi}}{2}\frac{\Gamma(1-\nu)}{\Gamma(\frac{3}{2}+\nu)}
\left(\frac{x_{min}p}{2}\right)^{2\nu}.
\end{equation}

\noindent Introducing the cutoff $\Lambda=2/x_{min}$ we finally have

\begin{equation}
\label{eqn:FAnyGaugeGrapheneFinal}
F(p,\xi)=
\frac{\sqrt{\pi}}{2}\frac{\Gamma(1-\nu)}{\Gamma(\frac{3}{2}+\nu)}
\left(\frac{p^{2}}{\Lambda^{2}}\right)^{\nu}.
\end{equation}

\noindent This is the nonperturbative form of the fermion propagator for graphene in any covariant gauge $\xi$. Its power-law behavior is consistent with the
multiplicative-renormalizable character of the theory. Notice that it is a different functional form of the corresponding transformation for the massless 
fermion propagator in QED$_3$ \eqn{eqn:SLKF3d}, and although we have derived it from the general expression in RQED$_{d_\gamma,d_e}$ \eqn{eqn:LKFTGraphene}, 
we could  have also defined it through the LKFT in ordinary QED in four space-time dimensions, $-\mi[\Delta_{4}(x_{min})-\Delta_{4}(x)]$, but integrating the
fermion momentum over a three-dimensional space-time. Proceeding in this form, we readily take into account the reduction of the power of $q$ in the 
denominator of the longitudinal part of the gauge boson propagator and redefinition of the gauge parameter~\cite{Gorbar:2001qt}, while retaining the gauge
covariance of the propagator itself. 



Since we eventually want to compare our full, nonperturbative, result \eqn{eqn:FAnyGaugeGrapheneFinal} with a perturbative evaluation of $F(p,\xi)$, we expand
\eqn{eqn:FAnyGaugeGrapheneFinal} in powers of $\alpha$:

\begin{equation}
\label{eqn:FAnyGaugeGrapheneWC}
F(p,\xi)=1+\frac{\xi\alpha}{4\pi}F_{1} 
+ \left(\frac{\xi\alpha}{4\pi}\right)^{2}F_{2}+\mathcal{O}(\alpha^{3}),
\end{equation}

\noindent with the expansion coefficients $F_{1}$ and $F_{2}$ given by


\begin{eqnarray}
\label{eqn:FAnyGaugeGrapheneWCF1Redef}
F_{1}&=& \ln\left(\frac{p^{2}}{\Lambda^{2}}\right) - \gamma_{E}-\psi\left(\frac{3}{2}\right) \nn \\
&=&\ln\left(\frac{p^{2}}{\Lambda^{2}}\right)+2\gamma_{E}+\ln(4)-2, \\
\label{eqn:FAnyGaugeGrapheneWCF2Redef}
F_{2}&=&\frac{1}{2}\left[\left(\ln\left(\frac{p^{2}}{\Lambda^{2}}\right) 
-\gamma_{E}-\psi(3/2)\right)^{2}-2\zeta(2)+4 \right] \nn \\
&=&\frac{1}{2}\left[\left(\ln\left(\frac{p^{2}}{\Lambda^{2}}\right)+2\gamma_{E}+\ln(4)
-2\right)^{2}-2\zeta(2)+4\right]\nonumber\\
\end{eqnarray}

\noindent where $\psi(z)$ is the digamma function, $\zeta(s)$ is the Riemann zeta function, and we have made use of the identity 
$\psi(3/2)=-\gamma_{E}-\ln(4) +2$.

In the next section we compare the expansion in \eqn{eqn:FAnyGaugeGrapheneWC}, with the coefficients shown in Eqs.~(\ref{eqn:FAnyGaugeGrapheneWCF1Redef}) 
and (\ref{eqn:FAnyGaugeGrapheneWCF2Redef}), against the one- and two-loop perturbative calculation of the fermion propagator.

\section{Perturbative constraints  of the fermion propagator in graphene}~\label{pert}

The fermion self-energy in RQED$_{d_\gamma,d_e}$ (and graphene in particular) has been calculated recently up to two loops in
Refs.~\cite{Teber:2012de,Kotikov:2013eha}. Our aim is to compare this perturbative calculation with a weak coupling expansion of our nonperturbative
LKFT result \eqn{eqn:FAnyGaugeGrapheneWC}. 


In RQED$_{d_{\gamma},d_{e}}$ the massless free fermion propagator is given by $S_{0}(p_{e})=\mi\cancel{p}/p^{2}$, where $p=(p_{0},\dots,p_{d_{e}-1})$ 
lies in the reduced fermion space, while the full fermion propagator is given by the solution of the Dyson equation
$S(p)=S_{0}(p)+S_{0}(p)\left(-\mi\Sigma(p)\right)S(p)$, where $\Sigma(p)$ is the fermion self-energy. The  general form of the solution of the Dyson 
equation for a massless fermion is $-\mi\cancel{p}S(p)=1/(1-\Sigma_{V}(p))$, where $\Sigma(p)=\cancel{p}\Sigma_{V}(p)$.
The vector part of the self-energy $\Sigma_{V}(p)$ is then related to the fermion wave function renormalization by

\begin{equation}
\label{eqn:SigmaAndFRel}
F(p;\xi)=\frac{1}{1-\Sigma_{V}(p;\xi)},
\end{equation}

\noindent where we have made explicit the gauge dependence of both quantities. As we mentioned above, $\Sigma_V(p;\xi)$ has been calculated
up to two loops for RQED$_{4,3}$. In the  $\overline{\text{MS}}$ regularization scheme, it is given by (see Ref.~\cite{Kotikov:2013eha})

\begin{multline}
\label{eqn:SPertRQED43}
\frac{1}{1-\Sigma_{V}(p;\xi)}= 1 + 
\frac{\alpha}{4\pi}\left[\frac{4}{9}-\frac{1-3\xi}{3}\overline{L}\right] \\
+ \left(\frac{\alpha}{4\pi}\right)^{2}
\left[\frac{(1-3\xi)^{2}}{18}(\overline{L}^{2}-2\zeta(2)+4)\right. \\
+ \left.4\frac{(3\xi+7)\overline{L}+48\zeta(2)}{27}
-8\zeta(2)(\overline{L}+2-\ln(4))-\frac{280}{27}\right],
\end{multline}

\noindent where 

\begin{equation}
\label{eqn:Lbar}
\overline{L}=\ln\left(-\frac{p^{2}}{\mu^{2}}\right)+\ln(4)-2,
\end{equation}

\noindent and $\mu$ is the renormalization mass scale. Note that the nontrivial terms in \eqn{eqn:SPertRQED43} contain a contribution 
that is proportional to the gauge parameter and one that does not vanish in Landau gauge.


We  now compare our LKFT result at weak coupling [Eqs.~(\ref{eqn:FAnyGaugeGrapheneWC}), (\ref{eqn:FAnyGaugeGrapheneWCF1Redef}), and
(\ref{eqn:FAnyGaugeGrapheneWCF2Redef})], to the perturbative calculation at one- and two-loop orders [\eqn{eqn:SPertRQED43}].
%
%
At ${\cal O}(\alpha)$, our LKFT result, \eqn{eqn:FAnyGaugeGrapheneWCF1Redef}, is proportional to the covariant gauge 
parameter---LKFT
gives only
terms of the type  $(\alpha\xi)^j$ when the starting point is the tree-level propagator. On the other hand, at this order, the perturbative result 
[\eqn{eqn:SPertRQED43}] has terms that are independent of the covariant gauge parameter. Since these terms cannot be obtained from a LKFT,
we should only compare terms that are proportional to $\alpha\xi$. Thus, $F_1$ defined in  \eqn{eqn:FAnyGaugeGrapheneWCF1Redef}  should be equivalent to 
$\overline{L}$ [\eqn{eqn:Lbar}]. This is indeed the case provided we identify

\begin{eqnarray}
\label{eqn:iden1}
\ln\left(\frac{p^{2}}{\Lambda ^{2}}\right)+2\gamma_{E}&\to& \ln\left(-\frac{p^{2}}{\mu^{2}}\right).
\end{eqnarray}
 
At ${\cal O}(\alpha^{2})$, the perturbative result has terms that are linear and quadratic in the covariant gauge parameter, apart from terms
that are independent of it. On the other hand, as can be seen from Eqs.~(\ref{eqn:FAnyGaugeGrapheneWC}) and (\ref{eqn:FAnyGaugeGrapheneWCF2Redef}), 
the LKFT only gives terms proportional to $\xi^{2}$ at order $\alpha^{2}$. This is expected given the structure of the LKFT. The terms linear in the 
covariant gauge parameter, to order $\alpha^{2}$, can only be recovered if we use a one-loop expression for $F(p;0)$ in \eqn{eqn:S0pSpace} 
[or equivalently in $X(x;0)$] as input into the LKFT~\cite{Bashir:2000}. This means that we should compare our $\alpha^{2}$ result, $F_{2}$, defined by [\eqn{eqn:FAnyGaugeGrapheneWCF2Redef}], only to the coefficient of $(\alpha\xi/(4\pi))^{2}$ in the perturbative result [\eqn{eqn:SPertRQED43}].  
Therefore, we see that

\begin{equation}
F_{2}\to \frac{9}{18}\left(\overline{L}^{2}-2\zeta(2)+4\right), 
\end{equation}

\noindent with the identification \eqn{eqn:iden1}, as expected.
Thus, we have shown that  there is full consistency between our LKFT result [\eqn{eqn:FAnyGaugeGrapheneWC}] and the perturbative
result [\eqn{eqn:SPertRQED43}] up to order $\alpha^{2}$. We hence predict the form of all the coefficients of the form $(\alpha\xi)^j$ in the all-order
perturbative expansion from our LKFT result~[\eqn{eqn:FAnyGaugeGrapheneFinal}].

\section{Conclusions}~\label{conclusions}

In this article, we have generalized the LKFT transformation for the fermion propagator in RQED$_{d_\gamma, d_e}$. The general transformation rule accounts 
for the integration of the bulk degrees of freedom of gauge bosons in the behavior of the longitudinal part of the corresponding reduced propagator and the
covariant gauge parameter. For the specific case of graphene, massless RQED$_{4,3}$, starting with the tree-level fermion propagator, we have obtained the 
full nonperturbative form of the fermion propagator in any covariant gauge. The power-law behavior of the wave function renormalization is in agreement with 
the multiplicative renormalizability features of the theory. We then confirmed that the weak coupling expansion of this propagator is in complete agreement 
with a perturbative calculation up to the two-loop level in terms of the form $(\alpha\xi)^2$.  We predict further agreement to higher
orders in $\alpha\xi$.

\begin{acknowledgements}
We acknowledge valuable discussions with Adnan Bashir along with support from CONACyT, CIC-UMSNH M\'exico, under Grants No.  CB-2014-242117, No. 4.22 and No. 522762. A.~A. acknowledges support from CONACyT M\'exico and Gomal University Pakistan. Y.~C.~S acknowledges support from PRODEP M\'exico under Grant No. 1255804.
\end{acknowledgements}

\appendix

\section*{Appendix: Auxiliary integrals}

Here we collect some useful integrals:

\begin{equation}
\label{eqn:ThetaMasterIntegral}
\int_{0}^{\pi}\dx{\theta}\sin^{2a}\theta\me^{\mi b\cos\theta}=
\sqrt{\pi}\Gamma\left(a+\frac{1}{2}\right)
\left(\frac{2}{b}\right)^{a}J_{a}(b),
\end{equation}

\noindent with $a>-1/2$, and

\begin{equation}
\label{eqn:tMasterIntegral}
\int_{0}^{\infty}\dx{t}t^{a}J_{b}(t)=2^{a}
\frac{\Gamma\left(\frac{1+a+b}{2}\right)}
{\Gamma\left(\frac{1-a+b}{2}\right)},
\end{equation}
 
\noindent with $a+b>-1,\,a<1/2$, where $J_{a}(z)$ is the Bessel function of the 
first kind. From \eqn{eqn:ThetaMasterIntegral} we can derive another result
that is useful too. Applying $-\mi\frac{\partial}{\partial b}$ to 
\eqn{eqn:ThetaMasterIntegral}, using 
$\frac{\partial}{\partial b}b^{-a}J_{a}(b)=-b^{-a}J_{a+1}(b)$, which can be 
obtained by using the identities $2\frac{\partial}{\partial z}J_{a}(z)=
J_{a-1}(z)-J_{a+1}(z)$ and $2a J_{a}(z)=z\left(J_{a-1}(z)+J_{a+1}(z)\right)$ we have

\begin{eqnarray}
\label{eqn:DThetaMasterIntegral}
\int_{0}^{\pi}\dx{\theta}\cos\theta\sin^{2a}\theta\me^{\mi b\cos\theta}&&\nonumber\\
&&\hspace{-20mm}=
\mi\sqrt{\pi}\Gamma\left(a+\frac{1}{2}\right)
\left(\frac{2}{b}\right)^{a}J_{a+1}(b),
\end{eqnarray}

\noindent with $\;a>-1/2$.


\end{document}